\documentclass{sigchi}

\toappear{Submitted for consideration for the PhD qualifying exam}

\usepackage{balance}  
\usepackage{graphicx} 
\usepackage{times}    
\usepackage{url}      

\makeatletter
\def\url@leostyle{%
  \@ifundefined{selectfont}{\def\UrlFont{\sf}}{\def\UrlFont{\small\bf\ttfamily}}}
\makeatother
\def\pprw{8.5in}
\def\pprh{11in}

\begin{document}

\title{Kinsey Reporter: Citizen Science for Sex Research}

\numberofauthors{6}
\author{
  \alignauthor Clayton A Davis\thanks{Contact author. Email: claydavi@indiana.edu}\\
    \affaddr{Center for Complex Networks and Systems Research, Indiana University School of Informatics and Computing}\\
    \affaddr{Bloomington, IN, USA}\\
  \alignauthor Julia Heiman\\
    \affaddr{The Kinsey Institute for Research in Sex, Gender, and Reproduction}\\
    \affaddr{Bloomington, IN, USA}\\  
  \alignauthor Erick Janssen\\
    \affaddr{Institute for Family and Sexuality Studies, KU Leuven}\\
    \affaddr{Leuven, Belgium}\\
  \alignauthor Stephanie Sanders\\
    \affaddr{The Kinsey Institute for Research in Sex, Gender, and Reproduction}\\
    \affaddr{Bloomington, IN, USA}\\ 
  \alignauthor Justin Garcia\\
    \affaddr{The Kinsey Institute for Research in Sex, Gender, and Reproduction}\\
    \affaddr{Bloomington, IN, USA}\\ 
  \alignauthor Filippo Menczer\\
    \affaddr{Center for Complex Networks and Systems Research, Indiana University School of Informatics and Computing}\\
    \affaddr{Bloomington, IN, USA}\\ 
}

\maketitle

\begin{abstract}
Kinsey Reporter is a global mobile app to share, explore, and visualize anonymous data about sex. Reports are submitted via smartphone, then visualized on a website or downloaded for offline analysis. 
In this paper we present the major features of the Kinsey Reporter citizen science platform designed to preserve the anonymity of its contributors, and preliminary data analyses that suggest questions for future research.
\end{abstract}

\section{Introduction}

The most popular sex applications available on the app marketplaces, e.g., Tinder, Grindr, and OKCupid, are great for finding sexual partners, but what happens during those and other sexual encounters? 
The original Kinsey Reports \cite{Kinsey1948, Kinsey1953} famously used face-to-face interviews to ascertain what really happens between the sheets. 
Adapting Kinsey's survey methodology to use web technology has provided sex researchers with additional methods for data gathering, but the increasing ubiquity of mobile computing devices presents new opportunities. Compared to traditional PCs, smartphone platforms provide location services and greater access to users in the developing world.


Kinsey Reporter (\url{kinseyreporter.org}) is a global mobile survey platform to share, explore, and visualize anonymous data about sex.
Reports are submitted via smartphone, then visualized on the website or downloaded for offline analysis.
``Citizen sex scientists'' submit reports, each consisting of one or more surveys, after participating in or observing sexual activity. 
Surveys cover topics such as flirting, sexual activity, unwanted experience, consumption of pornography, and hormonal birth control side effects.
One of the goals of the projects is to reveal how norms and behaviors surrounding these topics vary depending on geography, so location data is crucial.
For example, some research indicates that hormonal birth control can differently affect women living in disparate regions of the world \cite{Vitzthum2014}.
However the sensitive nature of these topics necessitates that reports be kept anonymous to protect both the participants and the researchers. The current implementation of Kinsey Reporter balances both of these demands by collecting no personally-identifying details and by anonymizing submission time and location with user-selectable resolution: city, province, or country.

Kinsey Reporter is a joint project of The Kinsey Institute for Research in Sex, Gender, and Reproduction and the Center for Complex Networks and Systems Research, both at Indiana University, Bloomington. 
Version 2 of Kinsey Reporter is currently available for iOS in the Apple App Store and for Android on Google Play. To date, about 
8,300 
reports have been tabulated from 
32 
different countries. The results can be visualized at \url{kinseyreporter.org/explore}.

In this paper we present the major features of the Kinsey Reporter platform, and preliminary data that suggest interesting research questions for future study.

\section{Motivation and Design}
This project aims to contribute to several different bodies of research surrounding collection and analysis of sensitive and confidential data.

\begin{figure}
\includegraphics[width=\linewidth]{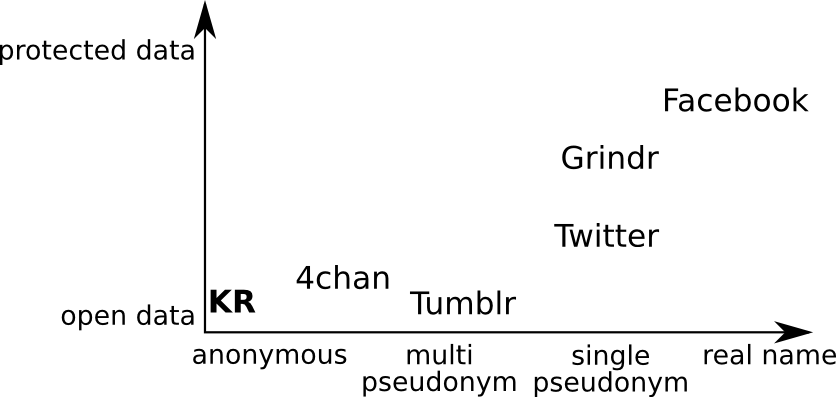}
\caption{Two axes of feature space with some social sites located on the axes for illustration. Placement is approximate, based on typical use of the sites. KR is Kinsey Reporter.}
\label{fig:anonymity-vs-openness}
\end{figure}

In order to delineate the various axes of feature space that it exists in, we have created Kinsey Reporter with an challenging set of constraints: an open-data, citizen-science platform that collects data on sensitive and private topics whilst protecting the anonymity of its users. To be clear, this represents somewhat of an extreme case as visualized in Figure~\ref{fig:anonymity-vs-openness} where KR is all the way in the corner. Other applications dealing with data of varied sensitivity will naturally occupy design space in between.

With this project, we hope to stimulate discussion on the pros, cons, and best practices of using citizen science platforms to collect sensitive data about private behaviors.
One such question of best practices comes from noting the distinction between personal data that is ``donated'', as in citizen science platforms, as opposed to data that is exchanged for a service, as in the case of Foursquare or other such location-enabled apps.
Are the app owners' obligations to the users different in these cases? 
Kinsey Reporter intentionally constrains its data collection in order to provide anonymity and protect both user and app provider from liability, but in doing so, does it also provide an example of the kind of transparency and protection from unintended use that should be expected of donated data?

Successfully interviewing subjects about sensitive topics -- as did our namesake Alfred Kinsey -- requires first achieving a certain level of interpersonal trust.
Trust is not a binary variable however; varying degrees of trust lead to people sharing different amounts and types of information, depending on context \cite{Anthony2007}.
In this light, Kinsey Reporter can be seen as a case study of how the overall platform design can contribute to the user's perception of trust.

\begin{figure}
\includegraphics[width=\linewidth]{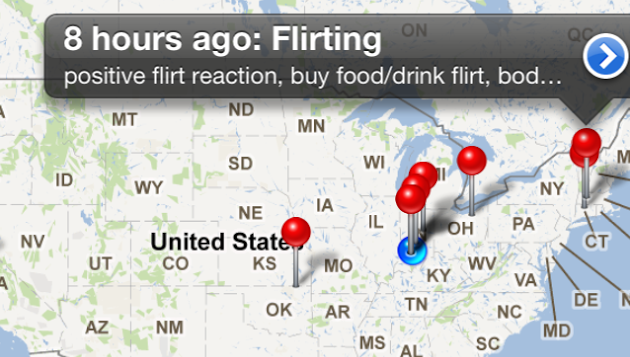}
 \caption{A cropped screenshot from the app, showing others' reports on a map.}
\label{fig:screenshot-map}
\end{figure}

\begin{figure}
\includegraphics[width=\linewidth]{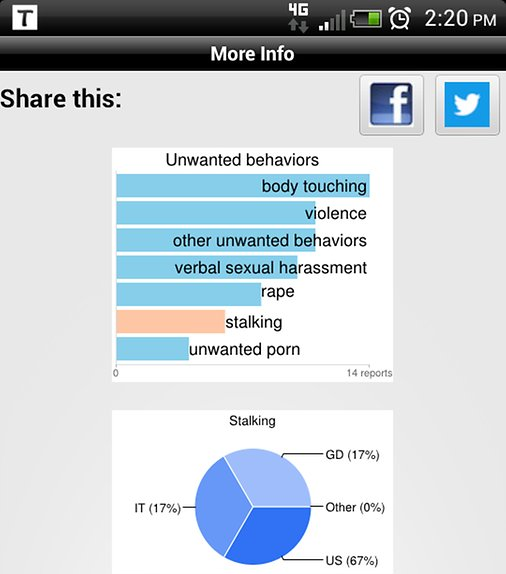}
 \caption{A screenshot from the app. After submission of a report, a user is able to explore how his or her submissions stack up to the aggregate. This user submitted the tag ``stalking'', thus it is highlighted in the aggregated result.}
\label{fig:screenshot-data}
\end{figure}

Research in medical contexts suggests that when people are unwilling to disclose sensitive, possibly stigmatizing or embarrassing information, their fear is along two axes: fear of judgement and fear that the information will leak beyond the intended bounds \cite{Stablein2012}. 
We hope that Kinsey Reporter's strict anonymity protections are understood by users to protect against both of these fears. In addition, our design acknowledges the social nature of trust and privacy \cite{Anthony2007, Benisch2011}. By providing users with feedback about others' responses in the form of maps (Fig.~\ref{fig:screenshot-map}) and aggregated survey results (Fig.~\ref{fig:screenshot-data}), we aim to employ subtle social encouragement to participate.

\section{Anonymity Features}

Survey topics can be highly sensitive. As such, neither the researcher nor any third-party with data access should be able to tie a report to a particular user.
This implies that we can not collect demographic or identifying information about particular users.

A stated goal of Kinsey Reporter as a citizen science project is to provide the data open to all.
As a result, pseudonymity (via usernames) becomes unacceptable: an attacker managing to de-anonymize one report should not be able to obtain the user's report history.
Thus we must abandon the concept of a user identity entirely, and design for total user anonymity.
These two axes of design space are illustrated in Figure~\ref{fig:anonymity-vs-openness}, along with some other social web sites.

\begin{figure}
\includegraphics[width=\linewidth]{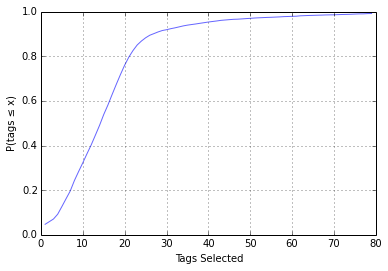}
\caption{Empirical cumulative distribution function (CDF) of the number of tags selected per report. The mean is approximately 16 tags per response. Only 1\% of reports have more than 80 tags.}
\label{fig:cdf-tags}
\end{figure}

Due to these hard requirements, we cannot analyze reports grouped by user. Empirically, users selected 136 thousand answers in 8300 reports submitted, yielding about 16 selections per report. The distribution of the number of tags per response can be seen in Fig.~\ref{fig:cdf-tags}. This makes it possible  to examine co-occurrence by report, especially among answers to questions belonging to the same survey. One can visualize these co-occurrence relationships with the \textit{Charts} tool at \url{kinseyreporter.org/explore}.

\subsection{Geo-temporal indistinguishability}

Suppose an attacker could observe a user interacting with Kinsey Reporter, for example in a coffee shop or other public place. If reports were posted immediately upon submission, the attacker could tie a report to the user by refreshing the report data until one matches the user's time and place of submission. To prevent this attack, reports are not made public immediately. Instead, recently-received reports sit in the database without timestamps and remain inaccessible to the public. These reports are then timestamped and made public only once a minimum number $k$ of reports are received from a particular geographic designation.  This protocol is akin to a user-adjustable, geo-temporal version of Sweeny's $k$-anonymity \cite{Sweeny2002}. The $k$ parameter is selected to tradeoff between data availability and user anonymity.

For a fixed $k$, another tradeoff exists between geographic and time resolution, as low-resolution geographic designations (e.g., country) will receive more reports in a given time than those with higher resolution (e.g., city). As a result, high geographic resolution corresponds to low temporal resolution, and vice-versa. This design is driven by an \emph{uncertainty principle}: it should not be possible to know both the location and the time of a report with high precision, as such information might reveal too much about an individual user. On the down-side, this protocol can make it difficult to compare data between rural and urban areas, due to the comparatively lower number of reports coming from a given rural municipality. It can also take a long time to accumulate $k$ reports from some of these more remote areas, causing reports to stay in ``limbo'' for long periods of time.

\subsection{No free text}

Free-text responses lend themselves to abuse. For instance, users could submit nonsense answers to skew results, as in Mountain Dew's infamous ``Dub the Dew'' campaign where the number one response was ``Hitler did nothing wrong'' \cite{Rosenfeld2012}. More importantly, free-text responses might be used to reveal (on purpose or not) someone's identity.
This deanonymization can be on purpose, by answering a question with text like ``John Doe committed sexual assault last night,'' or inadvertently, in the case of stylometric analysis. For example, textual analysis may be used to infer demographic information about the submitter such as age and gender, tie together multiple reports coming from the same submitter, and even possibly connect submissions to extant social media accounts \cite{Johansson2013, Korayem2013, Peersman2011}.

 For these reasons, users are not allowed to enter free-text responses to questions; all questions are multiple-choice. As a result, surveys must be very carefully crafted so as to avoid bias.

Unfortunately, there is little preventing users from ``stuffing the ballot box'' with spurious multiple-choice selections, since user anonymity precludes us from banning a user. This, however, cannot be done \textit{en-masse}, as described next. 

\subsection{Secure communication}

The app employs encryption and authentication in its communication with the Kinsey Reporter servers. First, public-key encryption (TLS/HTTPS) is a natural requirement to prevent eavesdropping on the network. Second, shared-key authentication is used to make it harder for an attacker script to submit a large volume of junk data to the Kinsey Reporter servers.

There are no trusted third parties. A geocoding service is used as a third-party to anonymize locations, but never receives sensitive data. The user's device only submits its location to the service, receiving the current city, state/province, and country. This lower-resolution location is then sent along with the report to Kinsey Reporter's servers.

In this case, the use of a third party for anonymizing location improves security. By ensuring that the exact location and the survey results are transmitted to separate remote servers, a hypothetical adversary needs to eavesdrop on two different encrypted connections in order to obtain unanonymized data.

\section{Survey Data}
\begin{figure}
\includegraphics[width=\linewidth]{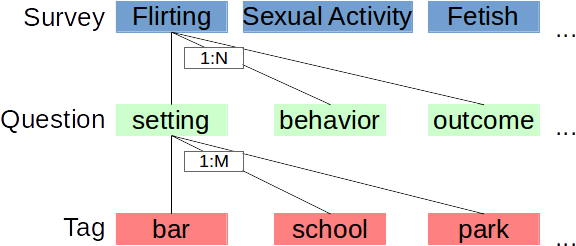}
\caption{Data model: surveys have associated questions, which in turn contain tags.}
\label{fig:structure}
\end{figure}
The user is presented with multiple surveys she can choose to answer, for example ``Sexual Activity'' or ``Flirting.'' Associated with each of these surveys are several questions, e.g., ``How many people involved?'' and ``Gender flirting.'' Likewise each question has a number of possible responses, called tags, such as ``multiple partners,'' or ``male flirting.'' Each of these relationships is one-to-many. The hierarchical data model is illustrated in Fig.~\ref{fig:structure}.

\subsection{Geography}

\begin{table}
\caption{Countries and US states with the most responses tabulated.}
\centering
\begin{tabular}{lrp{0.1in}|p{0.1in}lr}
\hline
\textbf{Country} & \textbf{Count} &&& \textbf{US State} & \textbf{Count} \\
\hline
USA & 7138 &&& Indiana    & 2907\\ 
Italy & 289 &&& California &  432\\ 
Canada & 237 &&& Texas      &  308\\ 
Netherlands & 172 &&& Oregon     &  249\\ 
Great Britain & 65 &&& Ohio       &  224\\ 
China & 63 &&& Michigan    &  199\\ 
Spain & 56 &&& Arizona   &  195\\ 
Turkey & 51 &&& Illinois   &  190\\ 
Denmark & 48 &&& Kentucky   &  159\\ 
Australia & 25 &&& New York    &  147\\ 
\hline
\end{tabular}
\label{tab:geo}
\end{table}

While reports have thus far been tabulated from 32 different countries, the vast majority come from the US. The top ten countries by number of responses can be seen in Table~\ref{tab:geo} along with the top ten US states. Among US states, Indiana is over-represented due to the disproportionate amount of media coverage Kinsey Reporter receives in its home state.

While Kinsey Reporter is explicitly global in its mission, it is limited by the current lack of localization --- the app and the surveys are currently only available in English. The effects of this limitation are observable in that four of the top five countries are largely English-speaking. Adding other languages will increase the potential user base of the platform and enable better international comparison of responses.

\subsection{Survey items}

\begin{table}
\caption{Surveys and number of tabulated responses for each. Note that the Porn and Valentine's Day surveys were released much later than the others.}
\begin{tabular}{lr}
\hline
\textbf{Survey} & \textbf{Count} \\
\hline
Sexual Activity                               & 6605 \\
Flirting                                      & 1161 \\
Public Display of Affection                   &  858 \\
Sexual Fetish                                 &  827 \\
Porn                                          &  528 \\
Female Hormonal Birth Control Use and Effects &  519 \\
Unwanted Experience                           &  306 \\
Valentine's Day                               &  196 \\
\hline
\end{tabular}
\label{tab:surveys}
\end{table}

As with the geography, the distribution of responses across surveys illustrated in Table~\ref{tab:surveys} is skewed, with the Sexual Activity survey receiving more responses than all other surveys combined. In some ways this is expected simply from demographics: with sexual activity broadly defined, this survey applies to the most people. On the other hand we suspect a large amount of self-selection happening in the survey-choosing process with the more ``fun'' surveys receiving more responses.

\begin{figure}
\includegraphics[width=\linewidth]{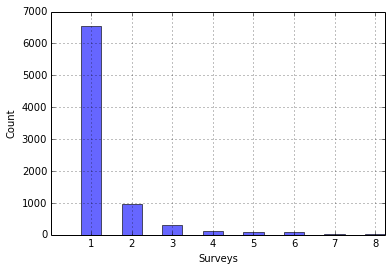}
\caption{Histogram of number of surveys responded to per report. The majority of reports contain answers to only a single survey.}
\label{fig:hist-surveys}
\end{figure}

With the inability to correlate user reports across responses, co-occurrence relationships must be analyzed from each response individually. Since users choose on average 16 tags per report, we have co-occurrence data that works well when correlating responses to a single survey, e.g., sexual activity given relationship status, as both questions are from the Sexual Activity survey. However, as shown in Fig.~\ref{fig:hist-surveys}, most responses contain answers from a single survey, thus somewhat limiting the ability to deduce co-occurrence relationships across surveys, e.g., sexual activity given flirting location.

\subsection{Preliminary results}

\begin{figure*}
\includegraphics[width=\linewidth]{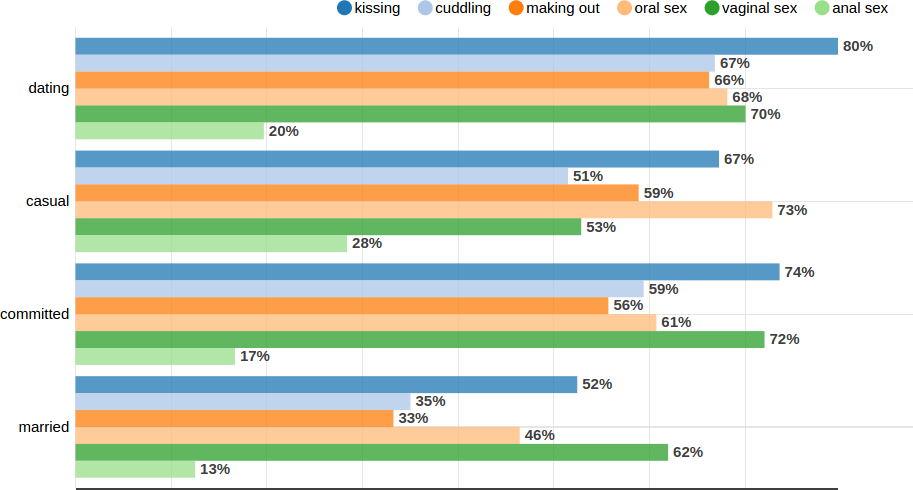}
 \caption{Sexual activity given relationship description, based on
 5453
 reports containing answers to both questions. Responses can include more than one activity. This chart was created with the Explore tool at \protect\url{kinseyreporter.org/explore}. }
\label{tab:activity-relationship}
\end{figure*}

While Kinsey Reporter is still a work in progress, it has tabulated a fair amount of data, albeit from a rather narrow set of surveys and locations. One co-occurrence relationship from the Sexual Activity survey is shown in Fig.~\ref{tab:activity-relationship}, illustrating sexual activity given relationship description. The chart shows expected relationships, such as increased incidence of oral sex amongst those in casual relationships \cite{Reiber2010},  
but also unexpected features such as the increased incidence of anal sex in casual encounters (392 reports). While these results are far from conclusive, these two examples combine to demonstrate the potential values of Kinsey Reporter's data: substantiating existing claims with a new source of data, and raising new questions to be further researched.

\section{Future Work}

Recent scholarly work has been done on privacy-preserving location-based services, including differential privacy models where the user selects the level of geographic resolution at which she is visible \cite{Andres2013}. As mentioned previously, Kinsey Reporter's current protocol can cause reports coming from low volume areas to stay in the queue for long periods of time. We are interested in adapting some of these new techniques to preserve geo-indistinguishability while reducing the time lag before making reports public.

\begin{figure}
\includegraphics[width=\linewidth]{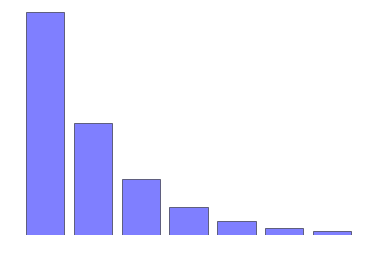}
\caption{A null model prediction for number of surveys responded to per report. The difference with Fig.~\ref{fig:hist-surveys} may suggest a design flaw.}
\label{fig:geometric-series}
\end{figure}

As mentioned previously and as seen in Fig.~\ref{fig:hist-surveys}, the fact that the vast majority of submissions contain answers to only one survey somewhat limits the ability to analyze inter-survey co-occurrence relationships.
A reasonable null model of user behavior might suggest half as many users submit responses to two surveys as opposed to a single submission, with the number of submissions decreasing geometrically with the number of surveys responded to, as seen in Figure \ref{fig:geometric-series}.
The overwhelming tendency to answer a single survey suggests that users may not appreciate that a report can contain responses to more than one survey (as is indeed desirable from the researchers' viewpoint).

Future collaboration with the human-computer interaction community may yield insights about how users interpret and experience the app, which in turn will allow us to analyze the data using stronger assumptions.
We also welcome suggestions about necessary improvements and high-priority enhancements to attract higher volumes of data world-wide.
Internationalization is also a high priority at this juncture. Our focus is explicitly international but our current batch of surveys are all in English.

\subsection{Acknowledgements}

We are grateful to Giorgio Elia, Muthu Chidambaram, Clayton Sheets, Magesh Vadivelu, and Nate Johnson for contributions to the development of the Kinsey Reporter app, and to Virginia J. Vitzthum for consultation on the surveys. Other contributors to the project are listed at \url{kinseyreporter.org/faq}.

\bibliography{kinseyreporter}{}
\bibliographystyle{plain}
\end{document}